\documentstyle[12pt,world_sci]{article}
\begin{document}
\title{STATUS OF THE STANDARD MODEL}
\author{JONATHAN L. ROSNER \\
{\em Enrico Fermi Institute and Department of Physics,
University of Chicago\\
5640 S. Ellis Ave., Chicago, IL 60637, USA}}
\maketitle
\setlength{\baselineskip}{2.6ex}

\vspace{-1.7in}
\rightline{EFI 94-38}
\rightline{hep-ph/9408349}
\rightline{August 1994}
\vspace{-0.5in}
\leftline{Presented at DPF 94 Meeting}
\leftline{Albuquerque, NM, Aug.~2 -- 6, 1994}
\leftline{Proceedings to be published by World Scientific}
\vspace{1.2in}

\begin{abstract}
\small
The standard model of electroweak interactions is reviewed, stressing the top
quark's impact on precision tests and on determination of parameters of the
Cabibbo-Kobayashi-Maskawa (CKM) matrix. Some opportunities for the study of CP
violation in the decays of $b$-flavored mesons are mentioned, and the
possibility of a new ``standard model'' sector involving neutrino masses is
discussed.
\end{abstract}

\section{Introduction}

Precision tests of the electroweak theory \cite{GWS} have reached a mature
stage since their beginnings more than twenty years ago.  We can now
successfully combine weak and electromagnetic interactions in a description
which also parametrizes CP violation through phases in the
Cabibbo-Kobayashi-Maskawa (CKM) \cite{cab,KM} matrix.  The mass quoted recently
by the CDF Collaboration \cite{CDFtop} for the top quark is one with which this
whole structure is quite comfortable.  Since this is the first DPF Meeting at
which we can celebrate the existence of more than a dozen top quark candidates
rather than just one or two, it is appropriate to review the impact of the top
quark's observation in the context of a wide range of other phenomena.  While
the evidence for the top quark could certainly benefit from a factor of four
greater statistics, it seems safe to say that the top is here to stay.  Looking
beyond it for the next aspects of ``standard model physics,'' we shall propose
that the study of neutrinos is a key element in this program.

We begin in Section 2 with a brief review of aspects of the top quark, covered
more fully in Mel Shochet's plenary talk \cite{Mel} and in parallel sessions
\cite{CDFDPF,D0DPF,SP}.  Section 3 is devoted to electroweak physics, while
Section 4
describes the present status of information about the CKM matrix. Some aspects
of the study of CP violation in $B$ decays are mentioned in Section 5.  We
devote Section 6 to a brief overview of neutrino masses and Section 7 to an
even briefer treatment of electroweak symmetry breaking. Section 8 concludes.

\section{The top quark}

\subsection{Cross section and mass}

The CDF Collaboration \cite{CDFtop,Mel,CDFDPF} has reported $m_t =
174 \pm 10 ^{+13}_{-12}$ GeV/$c^2$. The production cross section $\sigma(\bar p
p \to t \bar t + \ldots) = 13.9^{+6.1}_{-4.8}$ pb at $\sqrt{s} = 1.8$ TeV is on
the high side of the QCD prediction (3 to 10 pb, depending on $m_t$).  The
D0 Collaboration \cite{D0DPF} does not claim evidence for the top, but if its
seven candidates (with a background of $3.2 \pm 1.1$) are ascribed to top, the
cross section for a 174 GeV/$c^2$ top quark is about $7 \pm 5$ pb. A cross
section in excess of QCD predictions could be a signature for new strongly
interacting behavior in the electroweak symmetry breaking sector \cite{HP,EL}
or for the production of new quarks \cite{Barg}. As we shall see, the mass
quoted by CDF is just fine to account for loop effects in electroweak processes
(through $W$ and $Z$ self-energies) and in giving rise to $B^0 -\overline{B}^0$
and CP-violating $K^0 - \overline{K}^0$ mixing.

\begin{figure}
\vspace{4in}
\caption{Masses of quarks and leptons on a logarithmic scale.  Widths of bars
denote uncertainties in quark masses.}
\end{figure}

\subsection{Family structure.}
The top quark is the last quark to fit into a set of three families of quarks
and leptons, whose masses are shown in Fig.~1:
\begin{equation}
\left ( \begin{array}{c} u \\ d \end{array} \right) ~~;~~~
\left ( \begin{array}{c} c \\ s \end{array} \right) ~~;~~~
\left ( \begin{array}{c} t \\ b \end{array} \right) ~~;
\end{equation}
\begin{equation}
\left ( \begin{array}{c} \nu_e \\ e \end{array} \right) ~~;~~~
\left ( \begin{array}{c} \nu_\mu \\ \mu \end{array} \right) ~~;~~~
\left ( \begin{array}{c} \nu_\tau \\ \tau \end{array} \right) ~~.
\end{equation}
Only the $\nu_\tau$ has not yet been directly observed.  If there are any more
quarks and leptons, the pattern must change, since the width of the $Z$ implies
there are only three light neutrinos \cite{SO}.

The question everyone
asks, for which we have no answer is:  ``Why is the top so heavy?''  In Section
6 we shall return to this question in another form suggested by Fig.~1, namely:
``Why are the neutrinos so light?'' Althought the top quark is by far the
heaviest, its separation from the charmed quark (on a logarithmic scale) is no
more than the $c - u$ separation. (Amusing exercises on systematics of
quark mass ratios have been performed \cite{BJm,RR}.) The fractional errors on
the masses of the heavy quarks $t,~b,~c$ are actually smaller than those on the
masses of the light quarks $s,~d,~u$.

\section{Electroweak physics}

\subsection{Electroweak unification}

In contrast to the electromagnetic interaction (involving photon exchange), the
four-fermion form of the weak interaction is unsuitable for incorporation into
a theory which makes sense to higher orders in perturbation theory. Already in
the mid-1930's, Yukawa proposed a particle-exchange model of the weak
interactions.  At momentum transfers small compared with the mass $M_W$ of the
exchanged particle, one identifies
\begin{equation} \label{eqn:low}
\frac{G_F}{\sqrt{2}} = \frac{g^2}{8M_W^2} ~~~~~,
\end{equation}
where $G_F = 1.11639(2) \times 10^{-5}$ GeV$^{-2}$ is the Fermi coupling, and
$g$ is a dimensionless constant.

The simplest version of such a theory \cite{GWS} predicted not only the
existence of a charged $W^{\pm}$, but also a massive neutral boson $Z^0$, both
of which were discovered in 1983.  The exchange of a $Z^0$ implied the
existence of new weak charge-preserving interactions, identified a decade
earlier.

The theory involves the gauge group SU(2) $\times$ U(1), with respective
coupling constants $g$ and $g'$. Processes involving $Z^0$ exchange at low
momentum transfers can be characterized by a four-fermion interaction with
effective coupling
\begin{equation} \label{eqn:loz}
\frac{G_F}{\sqrt{2}} = \frac{g^2 +{g'}^2}{8M_Z^2} ~~~.
\end{equation}
The electric charge is related to $g$ and $g'$ by
\begin{equation}
e = g \sin \theta = g' /cos \theta~~~,
\end{equation}
where $\theta$ is the angle describing the mixtures of the neutral SU(2) boson
and U(1) boson in the physical photon and $Z^0$.  These relations can be
rearranged to yield
\begin{equation}
M_W^2 = \frac{\pi \alpha}{\sqrt{2} G_F \sin^2 \theta}~~~;
\end{equation}
\begin{equation}
M_Z^2 = \frac{\pi \alpha}{\sqrt{2} G_F \sin^2 \theta \cos^2 \theta} ~~~.
\end{equation}

Using the $Z$ mass measured at LEP \cite{SO} and a value of the electromagnetic
fine structure constant $\alpha(M_Z^2) \simeq 1/128$ evaluated at the
appropriate momentum scale, one obtains a value of $\theta$ and a consequent
prediction for the $W$ mass of about 80 GeV/$c^2$, which is not too bad.
However, one must be careful to define $\alpha$ properly (in one convention it
is more like 1/128.9) and to take all vertex and self-energy corrections into
account.  Crucial contributions are provided by top quarks in $W$ and $Z$
self-energy diagrams \cite{Tini}.  Eq.~(\ref{eqn:loz}) becomes
\begin{equation}
\frac{G_F}{\sqrt{2}} \hat \rho =
\frac{g^2 + {g'}^2}{8M_Z^2} ~~~,
\end{equation}
where
\begin{equation} \label{eqn:rho}
\hat \rho \simeq 1 + \frac{3G_F m_t^2}{8 \pi^2 \sqrt{2}} ~~~,
\end{equation}
so that
\begin{equation}
M_Z^2 = \frac{\pi \alpha}{\sqrt{2} G_F \hat \rho \sin^2 \theta \cos^2 \theta}
{}~~~.
\end{equation}

The angle $\theta$ and the mass of the $W$ now acquire implicit dependence on
the top quark mass.  The quadratic dependence of $\hat \rho$ on $m_t$ is a
consequence of the chiral nature of the $W$ and $Z$ couplings to quarks; no
such dependence occurs in the photon self-energy, which involves purely vector
couplings.  Small corrections to the right-hand sides of Eqs.~(\ref{eqn:low})
and (\ref{eqn:loz}), logarithmic in $m_t$, also arise.  We have ignored a QCD
correction \cite{RW} which replaces $m_t^2$ by approximately $0.9 m_t^2$ in
Eq.~(\ref{eqn:rho}).  Taking this into account would increase our quoted
$m_t$ values by about 5\%.

\subsection{The Higgs boson}

The electroweak theory requires the existence of something in addition to $W$'s
and a $Z$ in order to be self-consistent.  For example, $W^+ W^-$ scattering
would violate probability conservation (``unitarity'') at high energy unless a
spinless neutral boson $H$ (the ``Higgs boson'') existed below about 1 TeV
\cite{LQT}.  This particle has been searched for in electron-positron
collisions with negative results below $M_H = 64$ GeV/$c^2$ \cite{SO}.

A Higgs boson contributes to $W$ and $Z$ self-energies and hence to $\hat\rho$.
We can express the deviation of $\hat
\rho$ from its value at some nominal top quark
and Higgs boson masses $m_t = 175$ GeV/$c^2$ and $M_H = 300$ GeV/$c^2$ by means
of $\Delta \hat \rho = \alpha T$, where
\begin{equation}
T \simeq \frac{3}{16 \pi \sin^2 \theta} \left[ \frac{m_t^2 - (175
{}~{\rm GeV})^2}{M_W^2} \right] - \frac{3}{8 \pi \cos^2 \theta}
\ln \frac{M_H}{300~{\rm GeV}} ~~~.
\end{equation}
One can also expand $\sin^2 \theta$ about its nominal value $x_0 \simeq 0.232$
calculated for the above top and Higgs masses and the $Z$ mass observed at LEP.
The angle $\theta$, the $W$ mass, and all other electroweak observables now
are functions of both $m_t$ and $M_H$ in the standard model.  Additional small
corrections to the right-hand sides of (\ref{eqn:low}) and (\ref{eqn:loz})
arise which are logarithmic in $M_H$.

\subsection{Electroweak experiments}

Direct $W$ mass measurements over the past few years, in GeV/$c^2$, include
$79.92 \pm 0.39$ \cite{OldCDFW},
$80.35 \pm 0.37$ \cite{UA2W},
$80.37 \pm 0.23$ \cite{NewCDFW},
$79.86 \pm 0.26$ \cite{D0W},
with average $80.23 \pm 0.18$ \cite{Wavg}).
The ratio $R_\nu \equiv\sigma(\nu N \to \nu + \ldots)/\sigma(\nu N \to \mu^- +
\ldots)$ depends on $\hat \rho$ and $\sin^2 \theta$ in such a way that it, too,
provides information mainly on $M_W$.  The average of a CCFR Collaboration
result presented at this conference \cite{CCFR} and earlier measurements at
CERN by the CDHS and CHARM Collaborations \cite{CDHS,CHARM} imply $M_W = 80.27
\pm 0.26$ GeV/$c^2$.

A number of properties of the $Z$, as measured at LEP \cite{LEP} and SLC
\cite{SLC}, are relevant to precise electroweak tests.  Global fits to these
data have been presented by Steve Olsen at this conference \cite{SO}.  For our
discussion we use the following:
\begin{equation}
M_Z = 91.1888 \pm 0.0044~{\rm GeV}/c^2~~~,
\end{equation}
\begin{equation}
\Gamma_Z = 2.4974 \pm 0.0038~{\rm GeV}~~~,
\end{equation}
\begin{equation}
\sigma_h^0 = 41.49 \pm 0.12~{\rm nb}~~~({\rm hadron~production~cross~
 section})~~~,
\end{equation}
\begin{equation}
R_\ell \equiv \Gamma_{\rm hadrons}/\Gamma_{\rm leptons} = 20.795 \pm 0.040
{}~~~,
\end{equation}
which may be combined to obtain the $Z$ leptonic width $\Gamma_{\ell\ell}(Z) =
83.96 \pm 0.18$ MeV.  Leptonic asymmetries include the forward-backward
asymmetry parameter $A_{FB}^\ell = 0.0170 \pm 0.0016$, leading to a value
\begin{equation}
\sin^2 \theta_\ell \equiv \sin^2 \theta_{\rm eff}  = 0.23107 \pm 0.0090~~~,
\end{equation}
and independent determinations of $\sin^2 \theta_{\rm eff} = (1/4)(1 -
[g_V^\ell/g_A^\ell]$ from the parameters
\begin{equation}
A_\tau \to \sin^2 \theta = 0.2320 \pm 0.0013~~~,
\end{equation}
\begin{equation}
A_e \to \sin^2 \theta = 0.2330 \pm 0.0014~~~.
\end{equation}
The last three values may be combined to yield
\begin{equation} \label{eqn:sqLEP}
\sin^2 \theta = 0.2317 \pm 0.0007~~~.
\end{equation}
We do not use values of $\sin^2 \theta$ from forward-backward asymmetries in
quark pair production, preferring to discuss them separately.  There have been
suggestions that the behavior of $Z \to b \bar b$ may be anomalous
\cite{Chiv,Tatsu}, while the asymmetries in charmed pair production still have
little statistical weight and those in light-quark pair production are subject
to some model-dependence.

The result of Eq.~(\ref{eqn:sqLEP}) may be compared with that based on the
left-right asymmetry $A_{LR}$ measured with polarized electrons at SLC
\cite{SLC}:
\begin{equation} \label{eqn:sqSLC}
\sin^2 \theta = 0.2294 \pm 0.0010~~~.
\end{equation}
The results are in conflict with one another at about the level of two standard
deviations.  This is not a significant discrepancy but we shall use the
difference to illustrate the danger of drawing premature conclusions about the
impact of electroweak measurements on the Higgs boson sector.

\subsection{Dependence of $M_W$ on $m_t$}
We shall illustrate the impact of various electroweak measurements by plotting
contours in the $M_W$ vs. $m_t$ plane \cite{JRRMP}.  A more general language
\cite{PT} is
better for visualizing deviations from the standard model, but space and time
limitations prevent its use here. As mentioned, QCD corrections to Eq.~(9) are
neglected.

The measurements of $M_W$ via direct observation and via deep inelastic
neutrino scattering, together with the CDF top quark mass, are shown as the
plotted points in Fig.~2.  The results are not yet accurate enough to tell us
about the Higgs boson mass, but certainly are consistent with theory.  We next
ask what information other types of measurements can provide.

The dependence of $\sin^2 \theta_{\rm eff}$ on $m_t$ and $M_H$ leads to the
contours of $\sin^2 \hat{\theta} \approx \sin^2 \theta_{\rm eff} - 0.0003$
shown in Fig.~3.  Here  $\sin^2 \hat{\theta}$ is a quantity defined \cite{GS}
in the $\overline{MS}$ subtraction scheme.  Also shown are bands corresponding
to the LEP and SLC averages (\ref{eqn:sqLEP}) and (\ref{eqn:sqSLC}).  Taken by
itself, the SLC result prefers a high top quark mass.  When combined with
information on the $W$ mass, however, the main effect of the SLC data is to
prefer a lighter Higgs boson mass (indeed, lighter than that already excluded
by experiments at LEP).

\begin{figure}
\vspace{4.5in}
\caption{Dependence of $W$ mass on top quark mass for various values of Higgs
boson mass. Curves, from left to right: $M_H = 50,~100,~200,~500,~1000$
GeV/$c^2$.  Horizontal error bars on plotted points correspond to CDF
measurement of $m_t = 174 \pm 17$ GeV/$c^2$.  Square:  average of direct
measurements of $W$ mass; cross: average of determinations based on ratio of
neutral-current to charged-current deep inelastic scattering cross sections.}
\end{figure}

The observation of parity violation in atomic cesium \cite{CW}, together with
precise atomic physics calculations \cite{Csrc}, leads to information on the
coherent vector coupling of the $Z$ to the cesium nucleus, encoded in the
quantity $Q_W = \hat \rho(Z - N - 4Z \sin^2 \theta)$. Contours of this quantity
are shown in Fig.~4.  The central value favored by experiment, $Q_W({\rm Cs}) =
-71.04 \pm 1.58 \pm 0.88$, lies beyond the upper left-hand corner of the
figure, but the present error is large enough to be consistent with
predictions.  Because of a fortuitous cancellation \cite{PGHS,MR}, this
quantity is very insensitive to standard-model parameters and very sensitive to
effects of new physics (such as exchange of an extra $Z$ boson).

\begin{figure}
\vspace{4.5in}
\caption{Dependence of $W$ mass on top quark mass for various values of Higgs
boson mass, together with contours of values of $\sin^2 \hat{\theta}
\approx \sin^2 \theta_{\rm eff} - 0.0003$ predicted
by electroweak theory (dot-dashed lines) and measured by LEP (lower region
bounded by dashed lines: 1 $\sigma$ limits) and SLD (upper region).}
\end{figure}

\subsection{Fits to electroweak observables}

We now present the results of a fit to the electroweak observables listed in
Table 1.  The ``nominal'' values (including \cite{DKS} $\sin^2
\theta_{\rm eff} = 0.2320$) are calculated for $m_t = 175$
GeV/$c^2$ and $M_H = 300$ GeV/$c^2$.  We use $\Gamma_{\ell \ell}(Z)$, even
though it is a derived quantity, because it has little correlation with other
variables in our fit.  It is mainly sensitive to the axial-vector coupling
$g_A^\ell$, while asymmetries are mainly sensitive to $g_V^\ell$.  We also omit
the total width $\Gamma_{\rm tot}(Z)$ from the fit, since it is highly
correlated with  $\Gamma_{\ell \ell}(Z)$ and mainly provides information on the
value of the strong fine-structure constant $\alpha_s$.  With $\alpha_s = 0.12
\pm 0.01$, the observed total $Z$ width is consistent with predictions.  The
partial width $\Gamma(Z \to b \bar b)$ will be treated separately below.

\begin{figure}
\vspace{4.5in}
\caption{Dependence of $W$ mass on top quark mass for various values of Higgs
boson mass, together with contours of values of weak charge $Q_W$ for cesium as
discussed in text.}
\end{figure}

\begin{table}
\begin{center}
\caption{Electroweak observables described in fit}
\medskip
\begin{tabular}{c c c c} \hline
Quantity        &   Experimental   &   Nominal    &  Experiment/     \\
                &      value       &    value     &   Nominal        \\ \hline
$Q_W$ (Cs)      & $-71.0 \pm 1.8^{~a)} $  &   $ -73.2^{~b)}$
   & $0.970 \pm 0.025$ \\
$M_W$ (GeV/$c^2$) & $80.24 \pm 0.15^{~c)}$  & $80.320^{~d)}$
   & $0.999 \pm 0.002$ \\
$\Gamma_{\ell\ell}(Z)$ (MeV) & $83.96 \pm 0.18^{~e)}$ & $83.90^{~f)}$
   & $1.001 \pm 0.002$ \\
$\sin^2 \theta_{\rm eff}$ & $0.2317 \pm 0.0007^{~f)}$ & $0.2320^{~g)}$
   & $0.999 \pm 0.003$ \\
$\sin^2 \theta_{\rm eff}$ & $0.2294 \pm 0.0010^{~h)}$ & $0.2320^{~g)}$
   & $0.989 \pm 0.004$ \\ \hline
\end{tabular}
\end{center}
\leftline{$^{a)}$ {\small Weak charge in cesium}\cite{CW}}
\leftline{$^{b)}$ {\small Calculation~\cite{MR} incorporating
atomic physics corrections}~\cite{Csrc}}
\leftline{$^{c)}$ {\small Average of direct measurements~\cite{Wavg}
and indirect information}}
\leftline{{\small \quad from neutral/charged current ratio in
deep inelastic neutrino scattering}\cite{CCFR,CDHS,CHARM}}
\leftline{$^{d)}$ {\small Including perturbative QCD corrections}~\cite{DKS}}
\leftline{$^{e)}$ {\small LEP average as of July, 1994}\cite{LEP}}
\leftline{$^{f)}$ {\small From asymmetries at LEP}\cite{LEP}}
\leftline{$^{g)}$ {\small As calculated \cite{DKS} with correction for
relation between $\sin^2 \theta_{\rm eff}$ and $\sin^2 \hat \theta$}\cite{GS}}
\leftline{$^{h)}$ {\small From left-right asymmetry in annihilations at SLC}
\cite{SLC}}
\end{table}

\begin{figure}
\vspace{3.5in}
\caption{Values of $\chi^2$ for fits to $m_t$ and to electroweak data described
in Table.  Solid curve:  full data set (5 d.~o.~f.); dashed curve:  without SLD
data (4 d.~o.~f.).}
\end{figure}

In addition to the variables in Table 1, we use the constraint
$m_t = 174 \pm 17$ GeV/$c^2$.  The results are shown in Fig.~5.  To
illustrate the impact of the SLD value of $\sin^2 \theta$, we show the
effect of omitting it.  Conclusions about the Higgs boson mass clearly are
premature, especially if they are so sensitive to one input.

\subsection{The decay $Z \to b \bar b$}
The ratio $R_b \equiv \Gamma(Z \to b \bar b)/\Gamma(Z \to {\rm hadrons})$ has
been measured to be slightly above the standard model prediction.  In view of
the extensive discussion of this process elsewhere at this conference
\cite{SO,Chiv,Tatsu}, we shall be brief.

If one allows $R_b$ and the corresponding quantity for charm, $R_c \equiv
\Gamma(Z \to c \bar c)/\Gamma(Z \to {\rm hadrons})$, to be free parameters in a
combined fit, the results are \cite{LEPHF}
\begin{equation}
R_b = 0.2202 \pm 0.0020~~;~~~R_c = 0.1583 \pm 0.0098~~~,
\end{equation}
to be compared with the standard model predictions $R_b = 0.2156 \pm 0.0006$
\cite{AKG} and $R_c \approx 0.171$ \cite{LEPHF}.  If one constrains $R_c$ to
the standard model prediction, one finds instead $R_b = 0.2192 \pm 0.0018$. The
discrepancy is at a level of about $2 \sigma$.

Predictions for $R_b$ in the standard model and in two different
two-Higgs-doublet models \cite{AKG} are shown in Fig.~6.  With appropriate
choices of masses for neutral and charged Higgs bosons, it is possible to
reduce the discrepancy between theory and experiment without violating other
constraints on the Higgs sector.

\begin{figure}
\vspace{5in}
\caption{Dependence of $R_b \equiv \Gamma(Z \to b \bar b)/\Gamma(Z \to {\rm
hadrons})$ on top quark mass.  Solid curves:  predictions of Minimal Standard
Model (MSM) for $R_b$ and $R_d \equiv \Gamma(Z \to d \bar d)/\Gamma(Z \to {\rm
hadrons})$.
Dashed curves:  two-Higgs models described in text with $\tan \beta = 70$
(upper) and 1 (lower).  Data point:  recent LEP and CDF measurements of
$R_b$ and $m_{\rm top}$.}
\end{figure}

A curious item was reported \cite{RB} in one of the parallel sessions of this
conference.  The forward-backward asymmetries in heavy-quark production,
$A^{0,b}_{FB}$ and $A^{0,c}_{FB}$, have been measured both on the $Z$ peak and
2 GeV above and below it.  All quantities are in accord with standard model
expectations except for $A^{0,c}_{FB}$ at $M_Z - 2$ GeV.  Off-peak asymmetries
can be a hint of extra $Z$'s \cite{EZ}.

\section{The CKM Matrix}

\subsection{Definitions and magnitudes}

The CKM matrix for three families of quarks and leptons will have four
independent parameters no matter how it is represented. In a
parametrization \cite{wp} in which the rows of the CKM matrix are labelled by
$u,~c,~t$ and the columns by $d,~s,~b$, we may write
\begin{equation}
V = \left ( \begin{array}{c c c}
V_{ud} & V_{us} & V_{ub} \\
V_{cd} & V_{cs} & V_{cb} \\
V_{td} & V_{ts} & V_{tb}
\end{array} \right )
\approx \left [ \matrix{1 - \lambda^2 /2 & \lambda & A \lambda^3 ( \rho -
i \eta ) \cr
- \lambda & 1 - \lambda^2 /2 & A \lambda^2 \cr
A \lambda^3 ( 1 - \rho - i \eta ) & - A \lambda^2 & 1 \cr } \right ]~~~~~ .
\end{equation}
Note the phases in the elements $V_{ub}$ and $V_{td}$.  These phases allow the
standard $V - A$ interaction to generate CP violation as a higher-order weak
effect.

The four parameters are measured as follows:

\begin{enumerate}

\item The parameter $\lambda$ is measured by a comparison of strange particle
decays with muon decay and nuclear beta decay, leading to $\lambda \approx \sin
\theta \approx 0.22$, where $\theta$ is the Cabibbo \cite{cab} angle.

\item The dominant decays of $b$-flavored hadrons occur via the element $V_{cb}
= A \lambda^2$.  The lifetimes of these hadrons and their semileptonic
branching ratios then lead to an estimate $A = 0.79 \pm 0.06$.

\item The decays of $b$-flavored hadrons to charmless final states allow one to
measure the magnitude of the element $V_{ub}$ and thus to conclude that
$\sqrt{\rho^2 + \eta^2} = 0.36 \pm 0.09$.

\item The least certain quantity is the phase of $V_{ub}$:  Arg $(V_{ub}^*) =
\arctan(\eta/\rho)$.  We shall mention ways in which information on this
quantity may be improved, in part by indirect information associated with
contributions of higher-order diagrams involving the top quark.

\end{enumerate}

The unitarity of V and the fact that $V_{ud}$ and $V_{tb}$ are very close to 1
allows us to write $V_{ub}^* + V_{td} \simeq A \lambda^3$, or, dividing by a
common factor of $A \lambda^3$,
\begin{equation}
\rho + i \eta ~~ + ~~ (1 - \rho - i \eta) = 1~~~.
\end{equation}
The point $(\rho,\eta)$ thus describes in the complex plane one vertex of a
triangle whose other two vertices are $(0,0)$ and $(0,1)$.

\subsection{Indirect information}

Box diagrams involving the quarks with charge 2/3 are responsible for $B^0 -
\overline{B}^0$ and CP-violating $K^0 - \overline{K}^0$ mixing in the standard
model.  Since the top quark provides the dominant contribution, one obtains
mainly information on the phase and magnitude of $V_{td}$.

The evidence for $B^0 - \overline{B}^0$ mixing comes from ``wrong-sign''
leptons in $B$ meson semileptonic decays and from direct observation of
time-dependent oscillations \cite{TDB}.  The splitting $\Delta m$ between mass
eigenstates is proportional to $f_B^2 |V_{td}|^2$ times a function of $m_t$
which can now be considered reasonably well-known.  Here $f_B$ is the $B$ meson
decay constant, analogous to the pion decay constant $f_\pi = 132$ MeV. Given a
range of $f_B$ and the experimental average for $B$ mesons of $\Delta m/\Gamma
= 0.71 \pm 0.07$, we can then specify a range of $|V_{td}|$, which is
proportional to $|1 - \rho - i \eta|$.  We then obtain a band in the
$(\rho,\eta)$ plane bounded by two circles with center (1,0).

The parameter $\epsilon$ characterizing CP-violating $K^0 - \overline{K}^0$
mixing arises from an imaginary part in the mass matrix which is dominated by
top quark contributions in the loop, with small corrections from charm. In the
limit of complete top dominance one would have Im ${\cal M} \sim f_K^2$
Im($V_{td}^2) \sim \eta (1 - \rho)$, so that $\epsilon = (2.26 \pm 0.02) \times
10^{-3}$ would specify a hyperbola in the $(\rho,\eta)$ plane with focus (1,0).
 The effect of charm is to shift the focus to about (1.4,0).

\begin{figure}
\vspace{4in}
\caption{Region in the $(\rho,\eta)$ plane allowed by various constraints.
Dotted semicircles denote
central value and $\pm 1 \sigma$ limits implied by $|V_{ub}/V_{cb}| = 0.08 \pm
0.02$.  Circular arcs with centers at $(\rho,\eta) = (1,0)$ denote constraints
from $B - \overline{B}$ mixing, while hyperbolae describe region bounded by
constraints from CP-violating $K - \overline{K}$ mixing.}
\end{figure}

\subsection{Constraints on $\rho$ and $\eta$}

When one combines the indirect information from mixing with the constraint on
$(\rho^2 + \eta^2)^{1/2}$ arising from $|V_{ub}/V_{cb}|$, one obtains the
allowed region shown in Fig.~7.  Here, in addition to parameters mentioned
earlier, we have taken $|V_{cb}| = 0.038 \pm 0.003$, the vacuum-saturation
factor $B_K = 0.8 \pm 0.2$, and $\eta_B B_B = 0.6 \pm 0.1$, where $\eta_B$
refers to a QCD correction.  Standard QCD correction factors are taken in the
kaon system \cite{JRCP}.  We have also assumed $f_B = 180 \pm 30$ MeV, for
reasons to be described presently.

The center of the allowed region is near $(\rho,\eta) = (0,0.35)$, with values
of $\rho$ between $- 0.3$ and $0.3$ and values of $\eta$ between 0.2 and 0.45
permitted at the $1 \sigma$ level.  The main error on the constraint from
$(\Delta m/\Gamma)_B$ arises from uncertainty in $f_B$, while the main error on
the hyperbolae associated with $\epsilon$ comes from uncertainty in the
parameter $A$, which was derived from $V_{cb}$.  Other sources of error have
been tabulated by Stone at this conference \cite{SSDPF}.

\subsection{Improved tests}

We can look forward to a number of sources of improved information about CKM
matrix elements ~\cite{JRCKM}.

{\it 4.4.1 Decay constant information} on $f_B$ affects the determination
of $|V_{td}|$ (and hence $\rho$) via $B^0 - \overline{B}^0$ mixing.  Lattice
gauge theories have become more bold in predicting heavy meson decay constants.
For example, one recent calculation obtains the values \cite{BLS}
$$
f_B = 187 \pm 10 \pm 34 \pm 15~~{\rm MeV}~~~,
$$
$$
f_{B_s} = 207 \pm 9 \pm 34 \pm 22~~{\rm MeV}~~~,
$$
$$
f_D = 208 \pm 9 \pm 35 \pm 12~~{\rm MeV}~~~,
$$
\begin{equation}
f_{D_s} = 230 \pm 7 \pm 30 \pm 18~~{\rm MeV}~~~,
\end{equation}
where the first errors are statistical, the
second are associated with fitting and lattice constant, and the third arise
from scaling from the static $(m_Q = \infty)$ limit.  An independent
lattice calculation \cite{DE} finds a similar value of $f_B$.
The spread between these and some other lattice
estimates \cite{LAT} is larger than the errors quoted above, however.

{\it Direct measurements} are available for the $D_s$ decay constant.
The WA75 collaboration \cite{WA75} has seen 6 -- 7 $D_s \to \mu \nu$ events and
conclude that $f_{D_s} = 232 \pm 69$ MeV.  The CLEO Collaboration
\cite{FDSCLEO} has a much larger statistical sample; the main errors arise from
background subtraction and overall normalization (which relies on the $D_s \to
\phi \pi$ branching ratio).  Using several methods to estimate this branching
ratio, Muheim and Stone \cite{MS} estimate $f_{D_s} = 315 \pm 45$ MeV.  We
average this with the WA75 value to obtain $f_{D_s} = 289 \pm 38$ MeV.  A
recent value from the BES Collaboration \cite{BES}, $f_{D_s} = 434 \pm 160$ MeV
(based on one candidate for $D_s \to \mu \nu$ and two for $D_s \to \tau \nu$),
and a reanalysis by F. Muheim \cite{FM} using the factorization hypothesis
\cite{MS}, $f_{D_s} = 310 \pm 37$ MeV, should be incorporated in
subsequent averages.

{\it Quark models} can provide estimates of decay constants and their
ratios. In a non-relativistic model \cite{ES}, the decay constant $f_M$ of a
heavy meson $M = Q \bar q$ with mass $M_M$ is related to the square of the $Q
\bar q$ wave function at the origin by $f_M^2 = 12 |\Psi(0)|^2 / M_M$.  The
ratios of squares of wave functions can be estimated from strong hyperfine
splittings between vector and pseudoscalar states, $\Delta M_{\rm hfs} \propto
|\Psi(0)|^2/m_Q m_q$.  The equality of the $D_s^* - D_s$ and $D^* - D$
splittings then suggests that
\begin{equation}
f_D/f_{D_s} \simeq (m_d/m_s)^{1/2} \simeq 0.8 \simeq f_B/f_{B_s}~~~,
\end{equation}
where we have assumed that similar dynamics govern the light quarks bound to
charmed and $b$ quarks.  Using our average for $f_{D_s}$, we find $f_D = (231
\pm 31)$ MeV\null.  One hopes that the Beijing Electron Synchrotron will be
able to find the decay $D \to \mu \nu$ via extended running at the $\Psi(3770)$
resonance, which was the method employed by the Mark III Collaboration to
obtain the upper limit \cite{MKIII} $f_D < 290$ MeV (90\% c.l.).

An absolute estimate of $|\Psi(0)|^2$ can been obtained using electromagnetic
hyperfine splittings \cite{AM}, which are probed by comparing isospin
splittings in vector and pseudoscalar mesons.  On this basis \cite{JRCP} we
estimate $f_B = (180 \pm 12)$ MeV. [This is the basis of the value taken above,
where we inflated the error arbitrarily.] We also obtain $f_{B_s} = (225 \pm
15)$ MeV from the ratio based on the quark model.

{\it 4.4.2 Rates and ratios} can constrain $|V_{ub}|$ and possibly
$|V_{td}|$. The partial width $\Gamma(B \to \ell \nu)$ is proportional to
$f_B^2 |V_{ub}|^2$.  The expected branching ratios are about $(1/2) \times
10^{-4}$ for $\tau \nu$ and $2 \times 10^{-7}$ for $\mu \nu$. Another
interesting ratio \cite{ALI} is $\Gamma(B \to \rho \gamma)/ \Gamma(B \to K^*
\gamma)$, which, aside from phase space corrections, should be
$|V_{td}/V_{ts}|^2 \simeq 1/20$.  At this conference, however, Soni \cite{AS}
has argued that there are likely to be long-distance corrections to this
relation.

{\it 4.4.3 The $K^+ \to \pi^+ \nu \bar \nu$ rate} is governed by loop
diagrams involving the cooperation of charmed and top quark contributions, and
lead to constraints which involve circles in the $(\rho, \eta)$ plane with
centers at approximately (1.4,0) \cite{LL}.  The favored branching ratio is
slightly above $10^{-10}$, give or take a factor of 2. A low value within this
range signifies $\rho > 0$, while a high value signifies $\rho < 0$. The
present upper limit \cite{LL} is $B(K^+ \to \pi^+ \nu \bar \nu) < 3
\times 10^{-9}$ (90\% c.l.).

{\it 4.4.4 The decays $K_L \to \pi^0 e^+ e^-$ and $K_L \to \pi^0 \mu^+
\mu^-$} are expected to be dominated by CP-violating contributions.  Two types
of CP-violating contributions are expected:  ``indirect,'' via the CP-positive
component $K_1$ component of $K_L = K_1 + \epsilon K_2$, and ``direct,'' whose
presence would be a detailed verification of the CKM theory of CP violation.
These are expected to be of comparable magnitude in most \cite{RVW,Dib} but not
all \cite{Ko} calculations, leading to overall branching ratios of order
$10^{-11}$.  The ``direct'' CP-violating contribution to $K_L \to \pi^0 \nu
\bar \nu$ is expected to be dominant, making this process an experimentally
challenging but theoretically clean source of information on the parameter
$\eta$ \cite{RVW}.

{\it 4.4.5 The ratio $\epsilon'/\epsilon$ for kaons} has long been viewed
as one of the most promising ways to disprove a ``superweak'' theory of CP
violation in neutral kaon decays \cite{RVW,sw}. The latest estimates
\cite{BEPS} are equivalent (for a top mass of about 170 GeV/$c^2$) to
$[\epsilon'/\epsilon]|_{\rm kaons} = (6 \pm 3) \times 10^{-4} \eta$, with an
additional factor of 2 uncertainty associated with hadronic matrix elements.
The Fermilab E731 Collaboration \cite{Gib} measures $\epsilon'/\epsilon = (7.4
\pm 6) \times 10^{-4}$, consistent with $\eta$ in
the range (0.2 to 0.45) we have already specified. The CERN NA31 Collaboration
\cite{NA31} finds $\epsilon'/\epsilon = (23.0 \pm 6.5) \times 10^{-4}$, which
is higher than theoretical expectations. Both groups are preparing new
experiments, for which results should be available around 1996.

{\it 4.4.6 $B_s - \overline{B}_s$ mixing} can probe the ratio $(\Delta
m)|_{B_s}/(\Delta m)|_{B_d} = (f_{B_s}/f_{B_d})^2 (B_{B_s}/B_{B_d})$
$|V_{ts}/V_{td}|^2$, which should be a very large number (of order 20 or more).
Thus, strange $B$'s should undergo many particle-antiparticle oscillations
before decaying.

The main uncertainty in an estimate of $x_s \equiv (\Delta m/ \Gamma)_{B_s}$ is
associated with $f_{B_s}$.  The CKM elements $V_{ts} \simeq -0.04$ and $V_{tb}
\simeq 1$ which govern the dominant (top quark) contribution to the mixing are
known reasonably well. We show in Table 2 the dependence of $x_s$ on $f_{B_s}$
and $m_t$. To measure $x_s$, one must study the time-dependence of decays to
specific final states and their charge-conjugates with resolution equal to a
small fraction of the $B_s$ lifetime (about 1.5 ps).

\begin{table}
\begin{center}
\caption{Dependence of mixing parameter $x_s$ on top quark mass and
$B_s$ decay constant.}
\medskip
\begin{tabular}{c c c c} \hline
\null \qquad $m_t$ (GeV/$c^2$)&  157  &  174  &  191  \\ \hline
$f_{B_s}$ (MeV)               &       &       &       \\
150                           &  7.6  &  8.9  & 10.2  \\
200                           & 13.5  & 15.8  & 18.2  \\
250                           & 21.1  & 24.7  & 28.4  \\ \hline
\end{tabular}
\end{center}
\end{table}

The question has been raised:  ``Can one tell whether $\eta \ne 0$ from $B_s -
\overline{B}_s$ mixing?''  The ratio of squares of decay constants for strange
and nonstrange $B$ mesons is expected to be $(f_{B_s}/f_{B_d})^2 \approx 1.19
\pm 0.1$ \cite{JRCP,ALBs}.  ALEPH claims \cite{ALBs}
\begin{equation}
\frac{\Delta m_s}{\Delta m_d} = (1.19 \pm 0.10) \left| \frac{V_{ts}}{V_{td}}
\right|^2 > 7.9~~~,
\end{equation}
leading to a bound $|1 - \rho - i \eta| < 1.84$. (An even more aggressive bound
equivalent to $|1 - \rho - i \eta| < 1.7$ was reported by V. Sharma \cite{VS}
in the plenary session.)  However, in order to show that the unitarity triangle
has nonzero area, assuming that $|V_{ub}/V_{cb}| > 0.27$, one must show $0.73 <
|1 - \rho - i \eta| < 1.27$.  With the above expression, taking the $B_s$ and
$B_d$ lifetimes to be equal, and assuming $0.64 < x_d < 0.78$, this will be so
if $13 < x_s < 27$.  An ``ideal'' measurement would thus be $x_s = 20 \pm 2$.

\section{CP violation and $B$ decays}

\subsection{Types of experiments}

Soon after the discovery of the $\Upsilon$ states it was realized that
CP-violating phenomena in decays of $B$ mesons were expected to be observable
and informative \cite{EGNR,BCP}.

{\it 5.1.1 Decays to CP non-eigenstates} can exhibit rate asymmetries
only if there are two different weak decay amplitudes and two different strong
phase shifts associated with them.  The weak phases change sign under charge
conjugation, while the strong phases do not. Thus, the rates for $B^+ \to K^+
\pi^0$ and $B^- \to K^- \pi^0$ can differ only if the strong phases differ in
the $I = 1/2$ and $I = 3/2$ channels, and interpretation of a rate asymmetry in
terms of weak phases requires knowing the difference of strong phases.  We
shall mention in Sec.~5.3 the results of a recent SU(3) analysis \cite{BPP}
which permits the separation of weak and strong phase shift information without
the necessary observation of a CP-violating decay rate asymmetry.

{\it 5.1.2 Decays of neutral $B$ mesons to CP eigenstates $f$} can
exhibit rate asymmetries (or time-dependent asymmetries) as a result of the
interference of the direct process $B^0 \to f$ and the two-step process $B^0
\to \bar B^0 \to f$ involving mixing.  Here one does not have to know the
strong phase shifts. Decay rate asymmetries directly proble angles of the
unitarity triangle.  One very promising comparison involves the decays $B^0 \to
J/\psi K_S$ and $\overline{B}^0 \to J/\psi K_S$, whose rate asymmetry is
sensitive to $\sin~[{\rm Arg}(V_{td}^2)] \equiv \sin(2 \beta)$. It is necessary
to know whether the decaying neutral $B$ meson was a $B^0$ or a
$\overline{B}^0$ at some reference time $t = 0$.  We now remark briefly on one
method \cite{GNR} for tagging such $B^0$ mesons using associated pions.

\subsection{$\pi - B$ correlations}

The correlation of a neutral $B$ meson with a charged pion is easily visualized
with the help of quark diagrams. By convention (the same as for kaons), a
neutral $B$ meson containing an initially produced $\bar b$ is a $B^0$.  It
also contains a $d$ quark.  The next charged pion down the fragmentation chain
must contain a $\bar d$, and hence must be a $\pi^+$.  Similarly, a $\bar B^0$
will be correlated with a $\pi^-$.

The same conclusion can be drawn by noting that a $B^0$ can resonate with a
positive pion to form an excited $B^+$, which we shall call $B^{**+}$ (to
distinguish it from the $B^*$, lying less than 50 MeV/$c^2$ above the $B$).
Similarly, a $\bar B^0$ can resonate with a negative pion to form a $B^{**-}$.
The combinations $B^0 \pi^-$ and $\bar B^0 \pi^+$ are {\it exotic},~i.e.,
they cannot be formed as quark-antiquark states.  No evidence for exotic
resonances exists.  Resonant behavior in the $\pi - B^{(*)}$ system, if
discovered, would be very helpful in reducing the combinatorial backgrounds
associated with this method.

\begin{figure}
\vspace{3.5in}
\caption{P-wave nonstrange resonances of a $c$ quark and a light ($\bar u$
or $\bar d$) antiquark.  Check marks with or without parentheses denote
observation of some or all predicted states.}
\end{figure}

The lightest states which can decay to $B \pi$ and/or $B^* \pi$ are P-wave
resonances of a $b$ quark and a $\bar u$ or $\bar d$. The expectations for
masses of these states  may be based on extrapolation from the known $D^{**}$
resonances, for which present data \cite{JB} and predictions \cite{EHQ} are
summarized in Fig.~8.

The 1S (singlet and triplet) charmed mesons have all been observed, while CLEO
\cite{JB} has presented at this conference evidence for all six (nonstrange and
strange) 1P states in which the light quarks' spins combine with the orbital
angular momentum to form a total light-quark angular momentum $j = 3/2$.  These
states have $J = 1$ and $J = 2$.  They are expected to be narrow in the limit
of heavy quark symmetry.  The strange 1P states are about 110 MeV heavier than
the nonstrange ones. In addition, there are expected to be much broader (and
probably lower) $j = 1/2~D^{**}$ resonances with $J = 0$  and $J = 1$.

For the corresponding $B^{**}$ states, one should add about 3.32 GeV (the
difference between $b$ and $c$ quark masses minus a small correction for
binding).  One then predicts \cite{EHQ} nonstrange $B^{**}$ states with $J =
(1,2)$ at (5755, 5767) MeV.  It is surprising that so much progess has been
made in identifying $D^{**}$'s without a corresponding glimmer of hope for the
$B^{**}$'s, especially since we know where to look.

\subsection{Decays to pairs of light pseudoscalars}

The decays $B \to (\pi \pi, \pi K, K \bar K)$ are a rich source of information
on both weak (CKM) and strong phases, if we are willing to use flavor SU(3)
symmetry.

The decays $B \to \pi \pi$ are governed by transitions $b \to d q \bar q~(q =
u,~d, \ldots)$ with $\Delta I = 1/2$ and $\Delta I = 3/2$, leading respectively
to final states with $I = 0$ and $I = 2$.  Since there is a single amplitude
for each final isospin but three different charge states in the decays, the
amplitudes obey a triangle relation:  $A(\pi^+ \pi^-) - \sqrt{2} A(\pi^0 \pi^0)
= \sqrt{2}  A(\pi^+ \pi^0)$.  The triangle may be compared with that for the
charge-conjugate processes and combined with information on time-dependent $B
\to \pi^+ \pi^-$ decays to obtain information on weak phases \cite{pipi}.

The decays $B \to \pi K$ are governed by transitions $b \to s q \bar q~(q =
u,~d, \ldots)$ with $\Delta I = 0$ and $\Delta I = 1$.  The $I = 1/2$ final
state can be reached by both $\Delta I = 0$ and $\Delta I = 1$ transitions,
while only $\Delta I = 1$ contributes to the $I = 3/2$ final state.
Consequently, there are three independent amplitudes for four decays, and one
quadrangle relation $A(\pi^+ K^0) + \sqrt{2}A(\pi^0 K^+) = A(\pi^- K^+) +
\sqrt{2} A(\pi^0 K^0)$.  As in the $\pi \pi$ case, this relation may be
compared with the charge-conjugate one and the time-dependence of decays to CP
eigenstates (in this case $\pi^0 K_S$) studied to obtain CKM phase information
\cite{pik}.

We re-examined \cite{BPP} SU(3) analyses \cite{OldSU} of the decays $B \to P P$
($P$ = light pseudoscalar). They imply a number of useful relations among $\pi
\pi,~\pi K$, and $K \bar K$ decays, among which is one relating $B^+$
amplitudes alone:
\begin{equation}
A(\pi^+ K^0) + \sqrt{2} A(\pi^0 K^+) = \tilde{r}_u \sqrt{2} A(\pi^+ \pi^0)~~~.
\end{equation}
Here $\tilde{r}_u \equiv (f_K/f_\pi)|V_{us}/V_{ud}|$.  This expression relates
one side of the $\pi \pi$ amplitude triangle to one of the diagonals of the
$\pi K$ amplitude quadrangle, and thus reduces the quadrangle effectively to
two triangles, simplifying previous analyses \cite{pik}.  Moreover, since one
expects the $\pi^+ K^0$ amplitude to be dominated by a penguin diagram (with
expected weak phase $\pi$) and the $\pi^+ \pi^0$ amplitude to have the phase
$\gamma = {\rm Arg}~V_{ub}^*$, the comparison of this last relation and the
corresponding one for charge-conjugate decays can provide information on the
weak phase $\gamma$.  We have estimated \cite{JRCP} that in order to measure
$\gamma$ to $10^{\circ}$ one needs a sample including about 100 events in the
channels $\pi^0 K^{\pm}$.

Further relations can be obtained \cite{BPP} by comparing the amplitude
triangles involving both charged and neutral $B$ decays to $\pi K$. By looking
at the amplitude triangles for these decays and their charge conjugates, one
can sort out a number of weak and strong phases.

Some combination of the decays $B^0 \to \pi^+ \pi^-$ and $B^0 \to \pi^- K^+$
has already been observed \cite{Battle}, and updated analyses in these and
other channels have been presented at this conference \cite{BPPC}.

\section{Neutrino masses and new mass scales}

\subsection{Expected ranges of parameters}

Referring back to Fig.~1 in which quark and lepton masses were displayed, we
see that the neutrino masses are at least as anomalous as the top quark mass.
There are suggestions that the known (direct) upper limits are far above the
actual masses, enhancing the puzzle.  Why are the neutrinos so light?

A possible answer \cite{seesaw} is that light neutrinos acquire Majorana masses
of order $m_M = m_D^2/M_M$, where $m_D$ is a typical Dirac mass and $M_M$ is a
large Majorana mass acquired by right-handed neutrinos. One explanation
\cite{SNU} of the apparent deficit of solar neutrinos as observed in various
terrestrial experiments invokes matter-induced $\nu_e \to \nu_\mu$ oscillations
in the Sun \cite{MSW} with a muon neutrino mass of a few times $10^{-3}$ eV.
With a Dirac mass of about 0.1 to 1 GeV characterizing the second quark and
lepton family, this would correspond to a right-handed Majorana mass $M_M =
10^9 - 10^{12}$ GeV.  As stressed by Georgi in his summary talk \cite{HG},
nobody really knows what Dirac mass to use for such a calculation, which only
enhances the value of experimental information on neutrino masses.  However,
using the above estimate, and taking a Dirac mass for the third neutrino
characteristic of the third quark and lepton family (in the range of 2 to 200
GeV), one is led by the ratios in Fig.~1 to expect the $\nu_\tau$ to be at
least a couple of hundred times as heavy as the $\nu_\mu$, and hence to be
heavier than 1 eV or so.  This begins to be a mass which the cosmologists could
use to explain at least part of the missing matter in the Universe \cite{AO}.

If $\nu_\mu \leftrightarrow \nu_\tau$ mixing is related to ratios of masses,
one might expect the mixing angle to be at least $m_\mu/m_\tau$, and hence
$\sin^2 2 \theta$ to exceed $10^{-2}$.

\subsection{Present limits and hints}

Some limits on neutrino masses and mixings have been summarized at a recent
Snowmass workshop \cite{MG}.  The E531
Collaboration \cite{Reay} has set limits for $\nu_\mu \to \nu_\tau$
oscillations corresponding to $\Delta m^2 < 1~{\rm eV}^2$ for large $\theta$
and $\sin^2 2 \theta <~{\rm (a~few)} \times 10^{-3}$ for large $\Delta m^2$.
The recent measurement of the zenith-angle dependence of the apparent
deficit in the ratio of atmospheric $\nu_\mu$ to $\nu_e$ -- induced events
in the Kamioka detector \cite{Kam,FH} can be interpreted in terms of neutrino
oscillations (either $\nu_\mu \to \nu_e$ or $\nu_\mu \to \nu_\tau$), with
$\Delta m^2$ of order $10^{-2}$ eV$^2$.
In either case maximal mixing, with $\theta = 45^{\circ}$, is the most highly
favored.  We know of at least one other case (the neutral kaon system) where
(nearly) maximal mixing occurs; perhaps this will serve as a hint to the
pattern not only of neutrino masses but other fermion masses as well.
However, it is not possible to fit the Kamioka atmospheric neutrino effect,
the apparent solar-neutrino deficit, and a cosmologically significant
$\nu_\tau$ using naive guesses for Dirac masses and a single see-saw scale.
Various schemes have been proposed involving near-degeneracies of two or more
neutrinos or employing multiple see-saw scales.


\subsection{Present and proposed experiments}

Opportunities exist and are starting to be realized for filling in a
substantial portion of the parameter space for
neutrino oscillations.  New short-baseline
experiments are already in progress at CERN \cite{CHORUS,NOMAD} and approved at
Fermilab \cite{E803}.  These are capable of pushing the $\nu_\mu
\leftrightarrow \nu_\tau$ mixing limits lower for mass differences $\Delta m^2$
of at least 1 eV$^2$.  New long-baseline experiments \cite{P822} would be
sensitive in the same mass range as the Kamioka result to smaller mixing
angles.  At this conference we have heard a preliminary result from a search
for $\bar \nu_\mu \to \bar \nu_e$ oscillations using $\bar \nu_\mu$ produced in
muon decays \cite{LSND}.  An excess of events is seen which, if interpreted in
terms of oscillations, would correspond to $\Delta m^2$ of several eV$^2$.  (No
evidence for oscillations was claimed.) A further look at the solar neutrino
problem will be provided by the Sudbury Neutrino Observatory \cite{SNO}.

We will not understand the pattern of fermion masses until we understand what
is going on with the neutrinos.  Fortunately this area stands to benefit from
much experimental effort in the next few years.

\subsection{Electroweak-strong unification}

Another potential window on an intermediate mass scale is provided by the
pattern of electroweak-strong unification.  If the strong and electroweak
coupling constants are evolved to high mass scales in accord with the
predictions of the renormalization group \cite{GQW}, as shown in Fig.~9(a),
they approach one another in the simplest SU(5) model \cite{GG}, but do not
really cross at the same point.  This ``astigmatism'' can be cured by invoking
supersymmetry \cite{Amaldi}, as illustrated in Fig.~9(b).  Here the cure is
effected not just by the contributions of superpartners, but by the richer
Higgs structure in supersymmetric theories.  The theory predicts many
superpartners below the TeV mass scale, some of which ought to be observable in
the next few years.

Alternatively, one can embed SU(5) in an SO(10) model \cite{SOten}, in which
each family of quarks and leptons (together with a right-handed neutrino for
each family) fits into a 16-dimensional spinor representation.  Fig.~9(c)
illustrates one scenario for breaking of SO(10) at two different scales, the
lower of which is a comfortable scale for the breaking of left-right symmetry
and the generation of right-handed neutrino Majorana masses.

\subsection{Baryogenesis}

The ratio of baryons to photons in our Universe is a few parts in $10^9$.
In 1967 Sakharov \cite{Sakh}
proposed three ingredients of any theory which sought to explain the
preponderance of baryons over antibaryons in our Universe:  (1) violation of C
and CP; (2) violation of baryon number, and (3) a period in which the Universe
was out of thermal equilibrium.  Thus our very existence may owe itself to CP
violation.  However, no consensus exists on a specific implementation of
Sakharov's suggestion.

\begin{figure}
\vspace{6in}
\caption{Behavior of coupling constants predicted by the renormalization group
in various grand unified theories. Error bars in plotted points denote
uncertainties in coupling constants measured at $M = M_Z$ (dashed vertical
line).  (a)  SU(5); (b) supersymmetric SU(5) with superpartners above 1 TeV
(dotted line) (c) example of an SO(10) model with an intermediate mass scale
(dot-dashed vertical line).}
\vglue 0.1cm
\end{figure}

A toy model illustrating Sakharov's idea can be constructed within an SU(5)
grand unified theory.  The gauge group SU(5) contains ``$X$'' bosons which can
decay both to $uu$ and to $e^+ \bar d$.  By CPT, the total decay rates of $X$
and $\bar X$ must be equal, but CP-violating rate differences $\Gamma(X \to uu)
\ne \Gamma(\bar X \to \bar u \bar u)$ and $\Gamma(X \to e^+ \bar d) \ne
\Gamma(\bar X \to e^- d)$ are permitted.  This example conserves $B - L$, where
$B$ is baryon number (1/3 for quarks) and $L$ is lepton number (1 for
electrons).

It was pointed out by 't Hooft \cite{tH} that the electroweak theory contains
an anomaly as a result of nonperturbative effects which conserve $B - L$ but
violate $B + L$.  If a theory leads to $B - L = 0$ but $B + L \ne 0$ at some
primordial temperature $T$, the anomaly can wipe out any $B+L$ as $T$ sinks
below the electroweak scale \cite{KRS}. Thus, the toy model mentioned above and
many others are unsuitable in practice.

\begin{figure}
\vspace{5.4in}
\caption{Mass scales associated with one scenario for baryogenesis.}
\end{figure}

One proposed solution is the generation of nonzero $B - L$ at a
high temperature, e.g., through the generation of nonzero lepton number $L$,
which is then reprocessed into nonzero baryon number by the `t Hooft anomaly
mechanism \cite{Yana}. We illustrate in Fig.~10 some aspects of the second
scenario.
The existence of a baryon asymmetry, when combined with information on
neutrinos, could provide a window to a new scale of particle physics.

Large Majorana masses acquired by
right-handed neutrinos would change lepton number by two units
and thus would be  ideal for generating a lepton asymmetry if Sakharov's other
two conditions are met.

The question of baryogenesis is thus shifted onto the leptons:  Do neutrinos
indeed have masses?  If so, what is their ``CKM matrix''?  Do the properties of
heavy Majorana right-handed neutrinos allow any new and interesting natural
mechanisms for violating CP at the same scale where lepton number is violated?
Majorana masses for right-handed neutrinos naturally violate left-right
symmetry and could be closely connected with the violation of $P$ and $C$ in
the weak interactions \cite{BKCP}.

An open question in this scenario, besides the precise form of CP violation at
the lepton-number-violating scale, is how this CP violation gets communicated
to the lower mass scale at which we see CKM phases.  Presumably this occurs
through higher-dimension operators which imitate the effect of Higgs boson
couplings to quarks and leptons.

\section{Electroweak symmetry breaking}

A key question facing the standard model of electroweak interactions is the
mechanism for breaking SU(2) $\times$ U(1).  We discuss two popular
alternatives; Nature may turn out to be cleverer than either.

\subsection{Fundamental Higgs boson(s)}

If there really exists a relatively light fundamental Higgs boson in the
context of a grand unified theory, one has to protect its mass from
large corrections.  Supersymmetry is the popular means for doing so.  Then
one expects a richer neutral Higgs structure, charged scalar bosons, and
superpartners, all below about 1 TeV.

\subsection{Strongly interacting Higgs sector}

The scattering of longitudinally polarized $W$ and $Z$ bosons becomes
strong and violates unitarity above a TeV or two if there does not
exist a Higgs boson below this energy \cite{LQT}.  The behavior is similar
to what one might expect for pion-pion scattering in the non-linear
sigma model above a few hundred MeV.  We wouldn't trust such a model above
that energy, and perhaps we should not trust the present version of
electroweak theory above a TeV.  If the theory really has a strongly
interacting sector, its $I = J = 0$ boson (like the $\sigma$ of QCD) may
be its least interesting and most elusive feature.  Consider, for example,
the rich spectrum of resonances in QCD, which we now understand in terms
of the interactions of quarks and gluons.  Such rich physics in electroweak
theory was a prime motivation for the construction of the SSC, and we wish
our European colleagues well in their exploration of this energy region
via the LHC.  [I am also indebted to T. Barklow \cite{TB} for reminding me
of the merits of TeV $e^+ e^-$ colliders in this regard.]

\section{Summary}

It appears that the top quark, reported by the CDF Collaboration at this
meeting, is here to stay.  We look forward to its confirmation by the D0
Collaboration and to more precise measurements of its mass and decay
properties.  Even now, its reported properties are in comfortable accord
with standard model expectations based on electroweak physics and mixing
effects.

Tests of the electroweak theory continue to achieve greater and greater
precision, with occasional excursions into the land of two- and three-standard
deviation discrepancies which stimulate our theoretical inventiveness but may
be no more than the expected statistical fluctuations. These effects include a
low value of $\sin^2 \theta$ from SLD, a high value of $R_b$ from the LEP
experiments, and an anomalous forward-backward asymmetry in charmed quark pair
production at an energy 2 GeV below the $Z$.

The Cabibbo-Kobayashi-Maskawa matrix provides an adequate framework
for explaining the observed CP violation, which is still confined to
a single parameter ($\epsilon$) in the neutral kaon system.  We have no
deep understanding of the origin of the magnitudes or phases in the CKM
matrix, any more than we understand the pattern of quark and lepton masses.
Nonetheless, there are many possibilities for testing the present picture,
a number of which involve rare kaon and $B$ meson decays.

Numerous opportunities exist for studying CP violation in $B$ decays, and
facilities are under construction for doing so.  In view of the widespread
attention given recently to asymmetric $B$ factories, I have menioned a couple
of alternatives which can be pursued at hadron machines and/or symmetric
electron-positron colliders.

Neutrino masses may provide us with our next ``standard'' physics.  I have
suggested that the mass scale of $10^9 - 10^{12}$ GeV is ripe for exploration
not only through the measurement of mass differences in the eV- and sub-eV
range, but also through studies of leptogenesis and partial unification of
gauge couplings. Searches for axions, which I did not mention, also can shed
some light on this mass window.

Alternatives for electroweak symmetry breaking, each with consequences for
TeV-scale physics, include fundamental Higgs bosons with masses protected by
supersymmetry, a strongly interacting Higgs sector, some new physics
which we may have thought of but not learned how to make tractable (like
compositeness of Higgs bosons, quarks, and leptons), or even something we
have not thought of at all.  We will really have solved the problem only
when we understand the bewildering question of fermion masses, a signal that
while the ``standard'' model may work very well, it is far from complete.

\section{Acknowledgements}

I am grateful to Jim Amundson, Aaron Grant, Michael Gronau, Oscar Hern\'andez,
Nahmin Horowitz, Mike Kelly, David London, Sheldon Stone, Tatsu Takeuchi, and
Mihir Worah for fruitful collaborations on some of the topics mentioned here,
to them and to I. Dunietz, M. Goodman, and G. Zapalac for discussions, and to
the many other speakers in the plenary sessions for sharing
their insights and plans so that we could coordinate our efforts.
This work was supported in part by the United States Department of Energy
under grant No. DE AC02 90ER40560.

\def \ap#1#2#3{{\it Ann. Phys. (N.Y.)} {\bf#1} (#3) #2}
\def \apny#1#2#3{{\it Ann.~Phys.~(N.Y.)} {\bf#1} (#3) #2}
\def \app#1#2#3{{\it Acta Physica Polonica} {\bf#1} (#3) #2}
\def \arnps#1#2#3{{\it Ann. Rev. Nucl. Part. Sci.} {\bf#1} (#3) #2}
\def \arns#1#2#3{{\it Ann. Rev. Nucl. Sci.} {\bf#1} (#3) #2}
\def \ba88{{\it Particles and Fields 3} (Proceedings of the 1988 Banff Summer
Institute on Particles and Fields), edited by A. N. Kamal and F. C. Khanna
(World Scientific, Singapore, 1989)}
\def \baphs#1#2#3{{\it Bull. Am. Phys. Soc.} {\bf#1} (#3) #2}
\def \be87{{\it Proceedings of the Workshop on High Sensitivity Beauty
Physics at Fermilab,} Fermilab, Nov. 11-14, 1987, edited by A. J. Slaughter,
N. Lockyer, and M. Schmidt (Fermilab, Batavia, IL, 1988)}
\def \cn{Collaboration}
\def \cp89{{\it CP Violation,} edited by C. Jarlskog (World Scientific,
Singapore, 1989)}
\def \dpf91{{\it The Vancouver Meeting - Particles and Fields '91}
(Division of Particles and Fields Meeting, American Physical Society,
Vancouver, Canada, Aug.~18-22, 1991), ed. by D. Axen, D. Bryman, and M. Comyn
(World Scientific, Singapore, 1992)}
\def \dpff{{\it The Fermilab Meeting - DPF 92} (Division of Particles and
Fields
Meeting, American Physical Society, Fermilab, 10 -- 14 November, 1992), ed. by
C. H. Albright \ite~(World Scientific, Singapore, 1993)}
\def \efi{Enrico Fermi Institute Report No.~}
\def \hb87{{\it Proceeding of the 1987 International Symposium on Lepton and
Photon Interactions at High Energies,} Hamburg, 1987, ed. by W. Bartel
and R. R\"uckl (Nucl.~Phys.~B, Proc. Suppl., vol. 3) (North-Holland,
Amsterdam, 1988)}
\def \ib{{\it ibid.}~}
\def \ibj#1#2#3{{\it ibid.} {\bf#1} (#3) #2}
\def \ijmpa#1#2#3{{\it Int.~J. Mod.~Phys.}~A {\bf#1} (#3) #2}
\def \ite{{\it et al.}}
\def \jpb#1#2#3{{\it J. Phys.} B {\bf#1} (#3) #2}
\def \jpg#1#2#3{{\it J. Phys.} G {\bf#1} (#3) #2}
\def \kdvs#1#2#3{{\it Kong.~Danske Vid.~Selsk., Matt-fys.~Medd.} {\bf #1}
(#3) No #2}
\def \ky85{{\it Proceedings of the International Symposium on Lepton and
Photon Interactions at High Energy,} Kyoto, Aug.~19-24, 1985, edited by M.
Konuma and K. Takahashi (Kyoto Univ., Kyoto, 1985)}
\def \lat90{{\it Results and Perspectives in Particle Physics} (Proceedings of
Les Rencontres de Physique de la Vallee d'Aoste [4th], La Thuile, Italy, Mar.
18-24, 1990), edited by M. Greco (Editions Fronti\`eres, Gif-Sur-Yvette,
France,
1991)}
\def \lg91{International Symposium on Lepton and Photon Interactions, Geneva,
Switzerland, July, 1991}
\def \lkl87{{\it Selected Topics in Electroweak Interactions} (Proceedings of
the Second Lake Louise Institute on New Frontiers in Particle Physics, 15 --
21 February, 1987), edited by J. M. Cameron \ite~(World Scientific, Singapore,
1987)}
\def \mpla #1#2#3{{\it Mod. Phys. Lett.} A {\bf#1} (#3) #2}
\def \nc#1#2#3{{\it Nuovo Cim.} {\bf#1} (#3) #2}
\def \np#1#2#3{{\it Nucl. Phys.} {\bf#1} (#3) #2}
\def \oxf65{{\it Proceedings of the Oxford International Conference on
Elementary Particles} 19/25 Sept.~1965, ed.~by T. R. Walsh (Chilton, Rutherford
High Energy Laboratory, 1966)}
\def \pisma#1#2#3#4{{\it Pis'ma Zh. Eksp. Teor. Fiz.} {\bf#1} (#3) #2 [{\it
JETP Lett.} {\bf#1} (#3) #4]}
\def \pl#1#2#3{{\it Phys. Lett.} {\bf#1} (#3) #2}
\def \plb#1#2#3{{\it Phys. Lett.} B {\bf#1} (#3) #2}
\def \ppnp#1#2#3{{\it Prog. Part. Nucl. Phys.} {\bf#1} (#3) #2}
\def \pr#1#2#3{{\it Phys. Rev.} {\bf#1} (#3) #2}
\def \prd#1#2#3{{\it Phys. Rev.} D {\bf#1} (#3) #2}
\def \prl#1#2#3{{\it Phys. Rev. Lett.} {\bf#1} (#3) #2}
\def \prp#1#2#3{{\it Phys. Rep.} {\bf#1} (#3) #2}
\def \ptp#1#2#3{{\it Prog. Theor. Phys.} {\bf#1} (#3) #2}
\def \rmp#1#2#3{{\it Rev. Mod. Phys.} {\bf#1} (#3) #2}
\def \si90{25th International Conference on High Energy Physics, Singapore,
Aug. 2-8, 1990, Proceedings edited by K. K. Phua and Y. Yamaguchi (World
Scientific, Teaneck, N. J., 1991)}
\def \slac75{{\it Proceedings of the 1975 International Symposium on
Lepton and Photon Interactions at High Energies,} Stanford University, Aug.
21-27, 1975, edited by W. T. Kirk (SLAC, Stanford, CA, 1975)}
\def \slc87{{\it Proceedings of the Salt Lake City Meeting} (Division of
Particles and Fields, American Physical Society, Salt Lake City, Utah, 1987),
ed. by C. DeTar and J. S. Ball (World Scientific, Singapore, 1987)}
\def \smass82{{\it Proceedings of the 1982 DPF Summer Study on Elementary
Particle Physics and Future Facilities}, Snowmass, Colorado, edited by R.
Donaldson, R. Gustafson, and F. Paige (World Scientific, Singapore, 1982)}
\def \smass90{{\it Research Directions for the Decade} (Proceedings of the
1990 DPF Snowmass Workshop), edited by E. L. Berger (World Scientific,
Singapore, 1991)}
\def \smassb{{\it Proceedings of the Workshop on $B$ Physics at Hadron
Accelerators}, Snowmass, Colorado, 21 June -- 2 July 1993, ed.~by P. McBride
and C. S. Mishra, Fermilab report FERMILAB-CONF-93/267 (Fermilab, Batavia, IL,
1993)}
\def \stone{{\it B Decays}, edited by S. Stone (World Scientific, Singapore,
1994)}
\def \tasi90{{\it Testing the Standard Model} (Proceedings of the 1990
Theoretical Advanced Study Institute in Elementary Particle Physics),
edited by M. Cveti\v{c} and P. Langacker (World Scientific, Singapore, 1991)}
\def \ufn#1#2#3#4#5#6{{\it Usp.~Fiz.~Nauk} {\bf#1} (#3) #2 [Sov.~Phys. -
Uspekhi {\bf#4} (#6) #5]}
\def \yaf#1#2#3#4{{\it Yad. Fiz.} {\bf#1} (#3) #2 [Sov. J. Nucl. Phys. {\bf #1}
 (#3) #4]}
\def \zhetf#1#2#3#4#5#6{{\it Zh. Eksp. Teor. Fiz.} {\bf #1} (#3) #2 [Sov.
Phys. - JETP {\bf #4} (#6) #5]}
\def \zhetfl#1#2#3#4{{\it Pis'ma Zh. Eksp. Teor. Fiz.} {\bf #1} (#3) #2 [JETP
Letters {\bf #1} (#3) #4]}
\def \zp#1#2#3{{\it Zeit. Phys.} {\bf#1} (#3) #2}
\def \zpc#1#2#3{{\it Zeit. Phys.} C {\bf#1} (#3) #2}

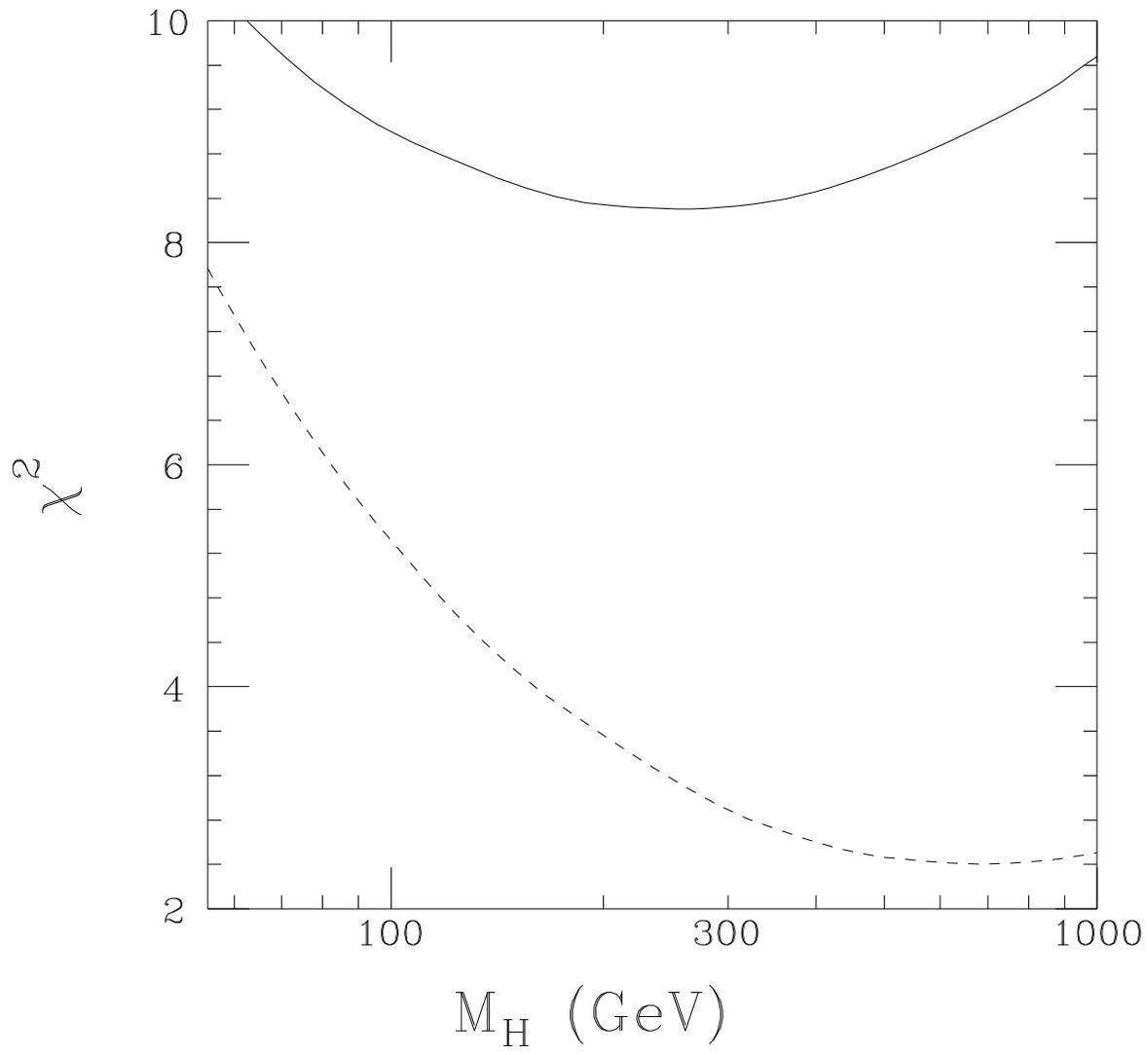
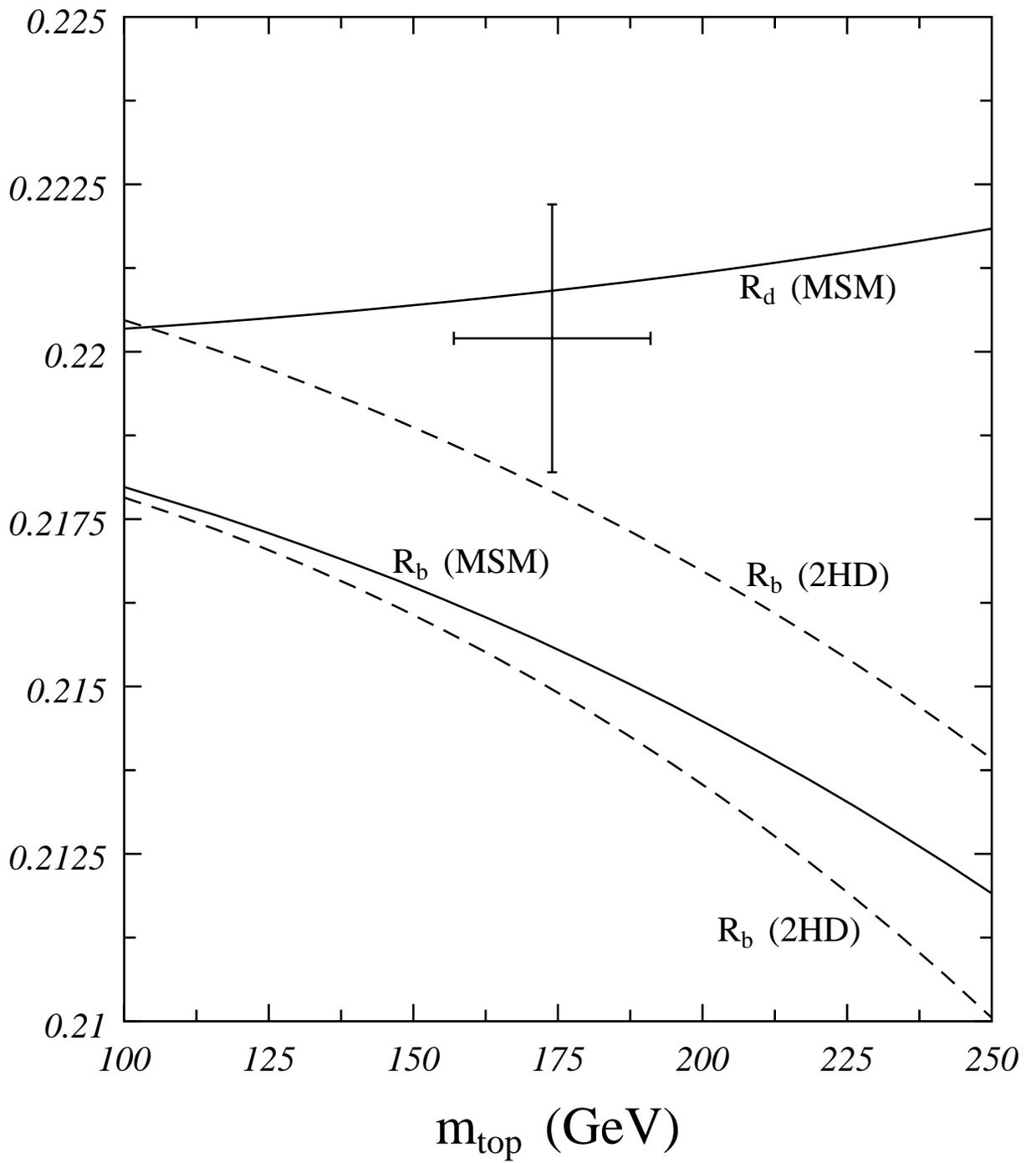
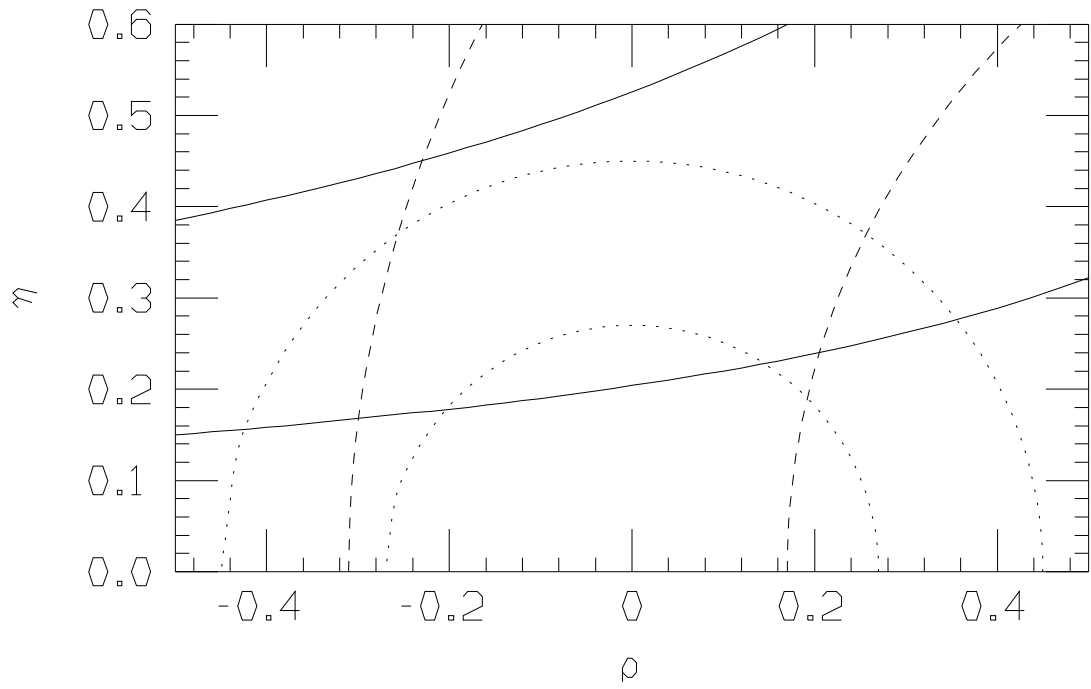
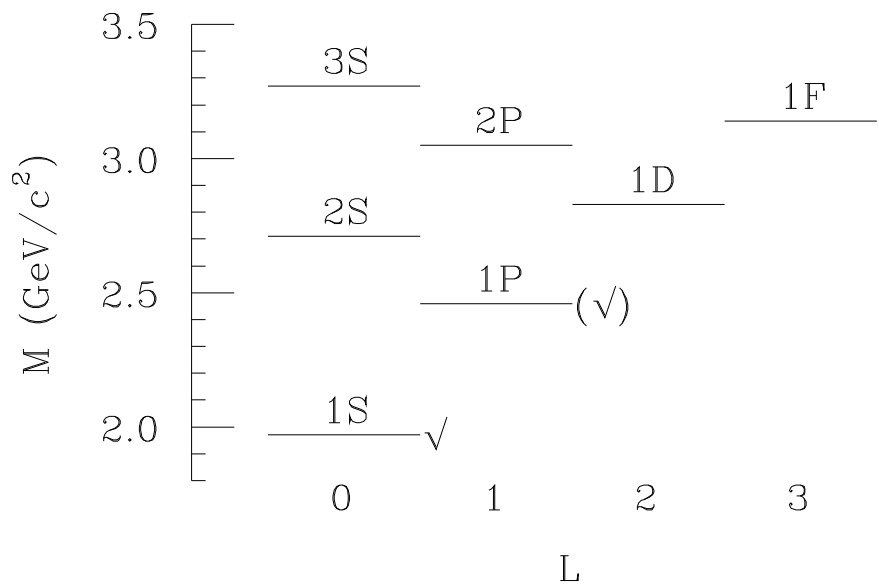
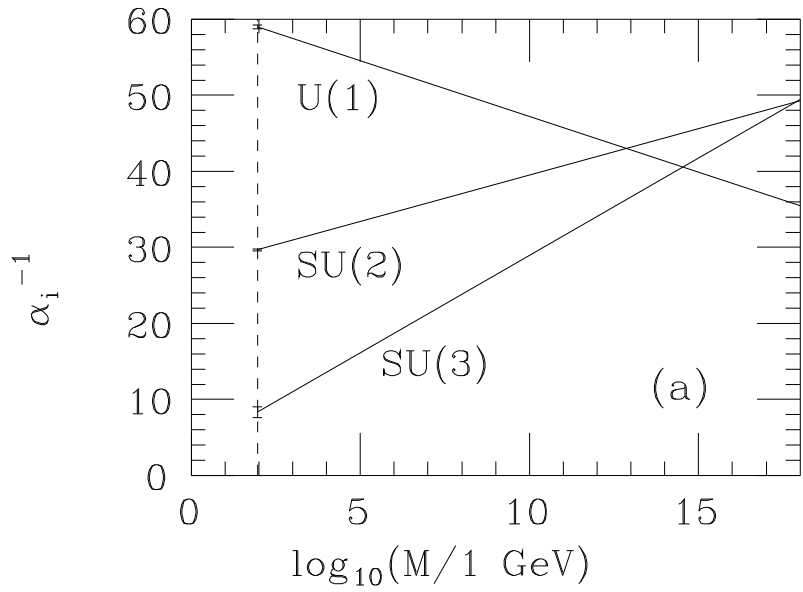
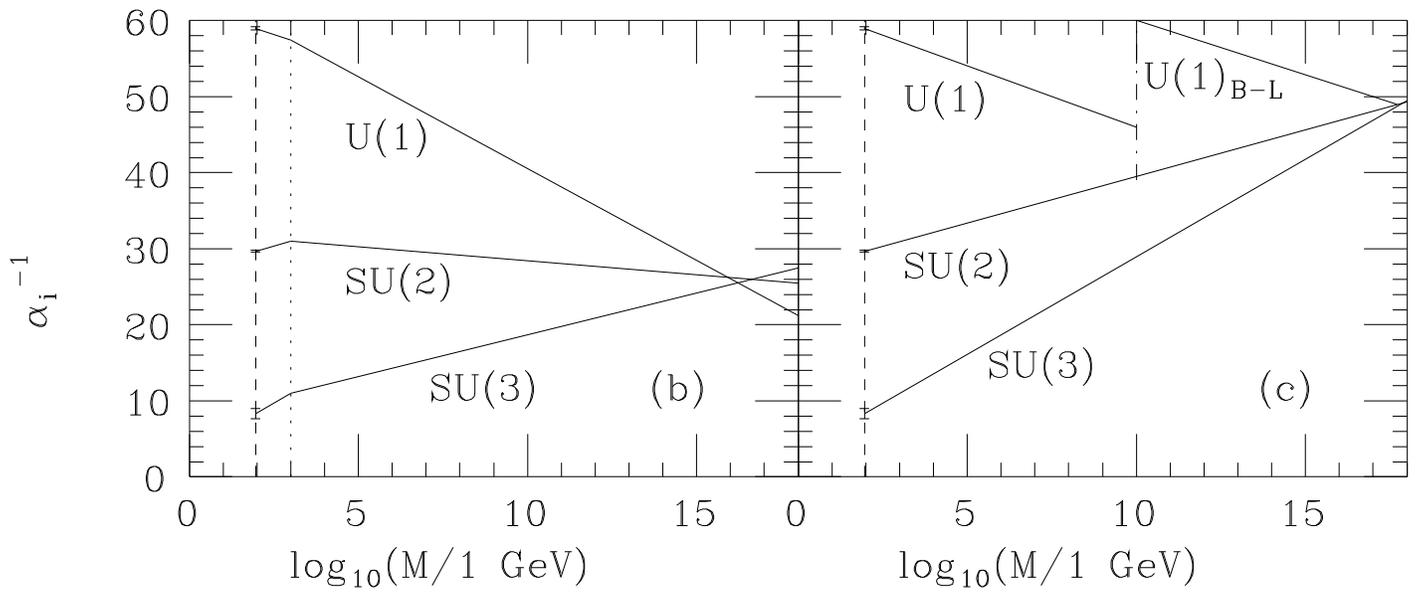
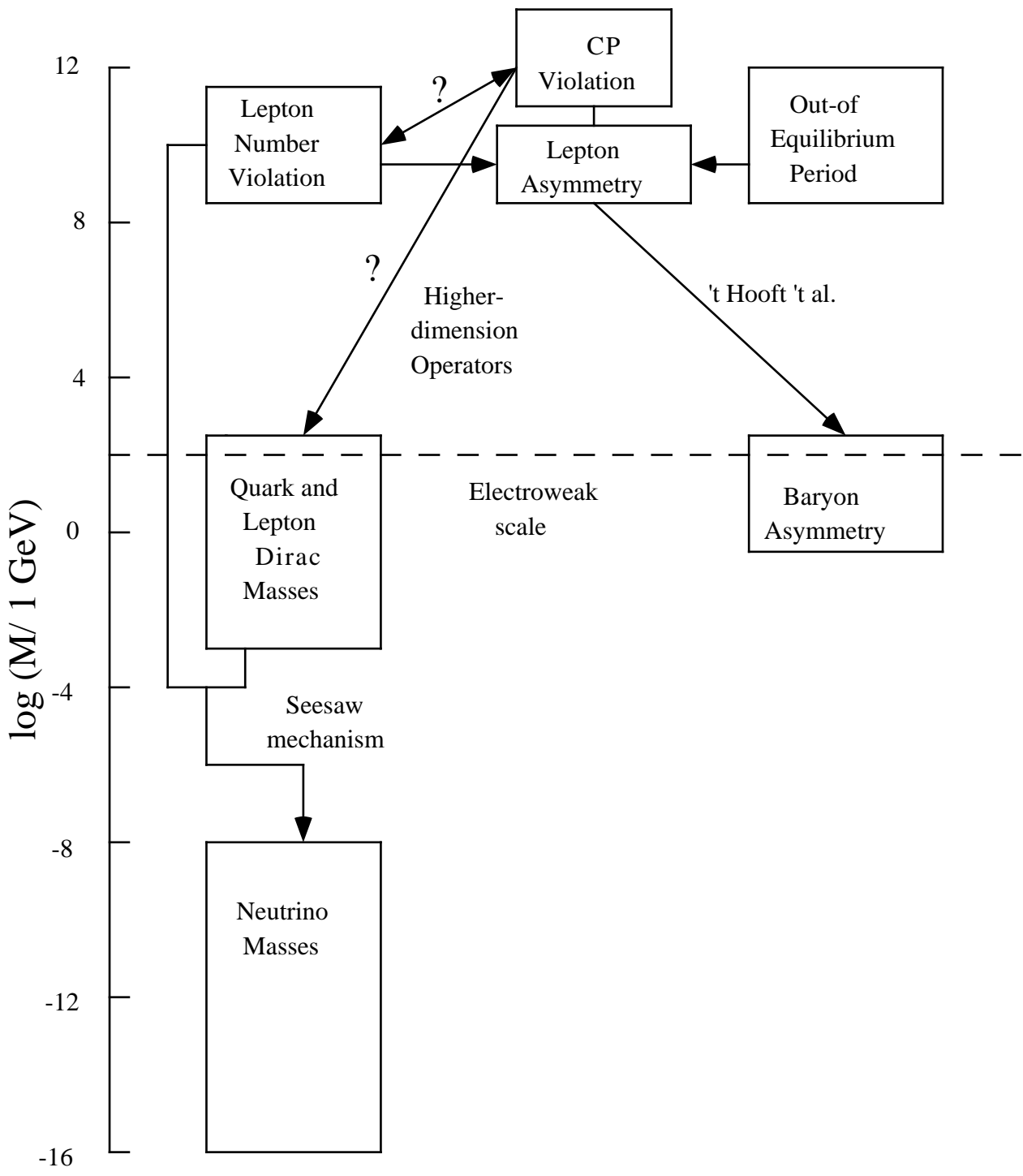
\end{document}